\documentclass[12pt]{article}
\usepackage{epsfig}
\usepackage{graphicx}
\usepackage{amsmath}
\usepackage{bm}

\begin{document}
\noindent
Chaos and Stochastic Models in Physics:
Ontic and Epistemic Aspects
\\
\\
\\
Sergio Caprara and Angelo Vulpiani
\\
\\
\small{Dipartimento di Fisica - Universit\`a Sapienza, Roma}
\\
\\
\\
{\bf Abstract} There is a persistent confusion about determinism and 
predictability. In spite of the opinions of some eminent philosophers 
(e.g., Popper), it is possible to understand that the two concepts are completely 
unrelated.  In few words we can say that determinism is ontic and has to do 
with how Nature behaves, while predictability is epistemic and is related 
to what the human beings are able to compute. An analysis of the Lyapunov 
exponents and the Kolmogorov-Sinai entropy shows how deterministic chaos, 
although with an epistemic character, is non subjective at all. This 
should clarify the role and content of stochastic models in the description
of the physical world.

\section{Introduction}

In the last decades scientists and philosophers showed an intense
interest for chaos, chance and predictability.
Some aspects of such topics are rather subtle, and in the literature
is not unusual to find wrong statements.
In particular it is important to avoid confusion on the fact that to be
deterministic (or stochastic)  is an ontic property of a system,
i.e. related to its own nature independently of our knowledge;
while predictability, and somehow chaos, have an epistemic character,
i.e. depend on our knowledge.
We will see  how the introduction of a probabilistic approach in
deterministic chaotic systems, although with an epistemic character, is 
not subjective.

Often in the past, the central goal of science has been though to be ``prediction and control'',
we can mention von Neumann's belief that powerful computers and a clever use of numerical analysis 
would eventually lead to accurate forecasts, and even to the control, of weather and climate:

{\it The computer will enable us to divide the atmosphere at any moment into stable regions and
unstable regions. Stable regions we can predict. Unstable regions we can control\footnote{Cited
by Dyson (2009).}}.

The great scientist von Neumann was wrong, but he did not know the phenomenon of deterministic chaos.

About half a century ago, thanks to the contribution of M. H\'enon, E. Lorenz and B. V. Chirikov (to cite 
just some of the most eminent scientists in the field), deterministic chaos was (re)discovered.
Such an event sure was scientifically important, e.g., as it clarifies topics like the different possible origins 
of the statistical laws and the intrinsic practical limits of the predictions. On the other hand, one has to 
admit that the term ``deterministic chaos'' can be seen as an oxymoron and induced the persistence of a certain 
confusion about concepts as determinism, predictability and stochastic laws. Our aim is to try to put some 
order into this matter, discussing some aspects of deterministic chaos which, in our opinion, are often 
misunderstood, leading to scientifically, as well as philosophically, questionable and confused claims.

In spite of the fact that it is quite evident that Maxwell, Duhem, and Poincar\'e understood in a clear 
way the distinction between determinism and chaos, in the recent literature one can find a large spectrum 
of wrong statements on the conceptual impact of deterministic chaos, see Campbell and Garnett (1882). 
For instance, Prigogine and Stengers
(1994) claim that {\it the notion of chaos leads us to rethink the notion of ``law of nature''}. In a book 
on statistical physics (Vauclair 1993), one can read that as consequence of chaos {\it the deterministic 
approach fails}. Sir James Lightill (1986) in a lecture to the Royal Society on the 300th anniversary of 
Newton's Principia shows how to confuse determinism and prediction: {\it We are all deeply conscious today 
that the enthusiasm of our forebears for the marvelous achievements of Newtonian mechanics led them to make 
generalization in this area of predictability, which indeed we may generally have tended to believe before 
1960, but which we now recognize were false. We collectively wish to apologize for having misled the generally 
educated public by spreading ideas about the determinism of systems satisfying Newton's laws of motion, that 
after 1960 were to be proved incorrect}.

Chaos presents both ontic and epistemic aspects\footnote{We shall see how determinism refers to ontic 
descriptions, while predictability (and, in some sense, chaos) has an epistemic nature.} which may generate 
confusion about the real conceptual relevance of chaos. We shall see that chaos allows us to unambiguously
introduce probabilistic concepts in a deterministic world. Such a possibility is not merely the consequence 
of our limited knowledge of the state of the system of interest. Indeed, in order to account for this limited 
knowledge, one usually relies on a coarse-grained description, which requires a probabilistic approach. We will
see that many important features of the dynamics do not depend on the scale $\epsilon$ of the graining, if 
it is fine enough. At the same time, many results for the $\epsilon \to 0$ limit do not apply to the cases 
with $\epsilon=0$. Therefore, the probabilistic description of chaotic systems reveals one more instance of 
singular limits.

\section{About determinism}
The word determinism has often been used in fields other than physics, such as psychology and sociology, causing 
some bewilderment. There have been some misunderstandings about the meaning of determinism, and because, at times,
determinism has been improperly associated with reductionism, mechanicism and 
predictability (Chibbaro, {\it  et al.} 2014), it seems to us 
that a brief review of the notion of determinism is not useless.

For example, unlike the majority of modern physicists and mathematicians, by deterministic system Popper (1992) 
means a system governed by a deterministic evolution law, whose evolution can be in principle predicted with 
arbitrary accuracy:

{\it Scientific determinism is the doctrine that the state of any closed physical system at any
future instant can be predicted.}

In other words, Popper confuses determinism and prediction.

On the contrary, Russell gives the following definition, which is in agreement with the present mathematical  terminology:

{\it A system is said to be ``deterministic'' when, given certain data $e_1, e_2, ... , e_n$ at times 
$t_1, t_2, ..., t_n$, respectively, concerning this system, if $E_t$ is the state of the system at any (later) time $t$, 
there is a functional relation of the form}
$$
E_t=f(e_1, t_1, e_2, t_2, ..., e_n, t_n) \,.
$$

In the definition of Russell practical prediction is not mentioned.
 
The confusion about determinism and predictability is not isolated, see, e.g., Stone (1989) and Boyd
(1972) who examine in great detail arguments about the widespread opinion that {\it human behavior is 
not deterministic because it is not predictable}. 

Determinism amounts to the metaphysical doctrine that same events always follow from same antecedents. But, 
as Maxwell had already pointed out in 1873, it is impossible to confirm this fact, because nobody has ever 
experienced the same situation twice:

{\it It is a metaphysical doctrine that from the same antecedents follow the same consequences.
No one can gainsay this. But it is not of much use in a world like this, in which the same
antecedents never again concur, and nothing ever happens twice ... The physical axiom which
has a somewhat similar aspect is ``that from like antecedents follow like consequences''. But
here we have passed ... from absolute accuracy to a more or less rough approximation.}

In these few lines, Maxwell touches on issues which will be later investigated,
and anticipates their solution. The issues are:

1. the impossibility of proving (or refuting) the deterministic character of the laws
of Nature;

2. the practical impossibility of making long-term predictions for a class of phenomena,
referred to here as chaotic, despite their deterministic nature.

After the development of quantum mechanics, many think that discussing the deterministic nature 
of the laws of physics is too academic an exercise to deserve serious consideration. For instance, in a 
speech motivated by the heated controversy on chaos and determinism between philosophers and scientists, 
Kampen (1991) bluntly said that the problem does not exist, as it is possible to show that:

{\it the ontological determinism \`a la Laplace can neither be proved nor disproved on the basis
of observations}\footnote{In brief, van KampenÕs argument is the following. Suppose the existence of a world A 
which is not deterministic and consider a second world B obtained from the first using the following deterministic 
rule: every event in B is the copy of an event occurred one million years earlier in A. Therefore, all the 
observers in B and their prototypes live the same experiences despite the different natures of the two worlds.}.

It is not difficult to realize that determinism and predictability constitute two quite distinct issues, and 
the former does not imply the latter. Roughly speaking, determinism can be traced back to a vision of the nature 
of causality and can be cast in mathematical terms, by saying that the laws of nature are expressed by ordinary 
(or partial) differential equations. However, as noted by Maxwell, the objectively ontological determinism of the 
laws of nature cannot be proven; but one might find it convenient to use deterministic descriptions. Moreover, 
even at a macroscopic level, many phenomena are chaotic and, in some sense, appear to be ``random''. On the other 
hand, the microscopic phenomena described by quantum mechanics, fall directly within a probabilistic framework. 
When referring to observable properties, they appear ontologically and epistemologically non-deterministic.

\section{Two explicit examples}
In order to clarify the concepts of determinism, predictability and chaos let us discuss two deterministic 
systems whose behaviors are rather different. They do not have particular own relevance,
their choice is motivated just for pedagogical reasons:
\\
{\bf Example A} The pendulum (of length $L$):
$$
{ d^2 \theta \over dt^2}= -{g \over L}\sin \theta \,.
\eqno(1)
$$
According to well known mathematical theorems on differential equations 
the following results hold:
\\
a)  the initial condition $(\theta(0), d\theta(0)/dt)$ determines in a unique way the state of the system
$(\theta(t), d\theta(t)/dt)$ at any time $t$, in other words the system is deterministic;
\\
b) the motion is periodic, i.e., there exists a time $T$ (depending on the initial conditions)
such that  
$$
\Big(\theta(t+T), {d\theta(t+T) \over dt} \Big)=\Big(\theta(t), {d\theta(t) \over dt} \Big) \,;
$$
\\
c) the  time evolution  can be expressed via a function $F(t, \theta(0), d\theta(0)/dt)$:
$$
\theta(t)= F\Big(t, \theta(0), {d\theta(0) \over dt} \Big)\,.
$$ 
The function $F$ can be explicitly written only if 
$\theta(0)$ and $d\theta(0)/dt$ are small (and, in such a case, $T=  2 \pi \sqrt{L /g}$ is a constant,
independent of the initial conditions); however, in the generic case, $F$ can be easily 
determined with the desired precision.
\\
\\
{\bf Example B} Bernoulli's shift:
$$
x_{t+1}= 2 x_t \,\,\, mod \,\, 1 \,.
\eqno(2)
$$
Where the operation $mod\,\, 1$ corresponds to taking the fractional part of a number, e.g., 
$1.473 \,\, mod\,\, 1=0.473$. It is easy to understand that the above system is deterministic: 
$x_0$ determines $x_1$, which determines $x_2$ and so on. Let us show that the above system is 
chaotic: a small error in the initial conditions doubles at every step. Suppose that $x_0$ 
is a real number in the interval $[0,1]$, it can be expressed by an infinite sequence of $0$ and 
$1$:
$$
x_0={a_1\over 2}+{a_2\over 4}+ ... +{a_n\over 2^n}+ ...,
$$
where every $a_n$ takes either the value $0$ or the value $1$. The above binary notation allows us 
to determine the time evolution by means of a very simple rule: at every step, one has just move the 
``binary point'' of the binary expansion of $x_0$ by one position to the right and eliminate the integer 
part. For example, take
$$
x_0=0.11010000101110101010101100.....
$$
$$
x_1=0.1010000101110101010101100.......
$$
$$
x_2=0.010000101110101010101100.........
$$
$$
x_3=0.10000101110101010101100...........
$$
and so on. In terms of the sequence $\{ a_1, a_2, ...\}$, it becomes quite clear how crucially the temporal 
evolution depends on the initial condition. Let us consider two initial conditions $x^{(1)}_0$ and $x^{(2)}_0$ 
such that $|x^{(1)}_0-x^{(2)}_0| < 2^{-M}$ for some arbitrary (large) integer number $M$, this means that 
$x^{(1)}_0$ and $x^{(2)}_0$ have the first $M$ binary digits identical, and they may differ only afterwards. 
The above discussion shows that the distance between the points increases rapidly: for $t <M$ one has an 
exponential growth of the distance between the two trajectories
$$
|x^{(1)}_t- x^{(2)}_t |\sim |x^{(1)}_0- x^{(2)}_0|\, 2^t \,.
$$
As soon as $t>M$, one can only conclude that $|x^{(1)}_t- x^{(2)}_t|< 1$. Our system is chaotic: even an 
arbitrarily small error in the initial conditions eventually dominates the dynamics of the system, making 
long-term prediction impossible.
\\
\\
From the above discussion we saw how in deterministic systems one can have 
the following possible cases (in decreasing order of predictability):
\\
I- Explicit possibility to determine the future (pendulum in the limit of small oscillations);
\\
II- Good control of the prediction, without an explicit solution (pendulum with large oscillations);
\\
III- Chaos and practical impossibility of predictability (Bernoulli's shift).

\subsection{About the ontic/epistemic character of chaos}

One should also beware of the possible confusion between ontic and epistemic descriptions, when studying 
the topic of chaos. Determinism simply means that: given the same initial state ${\bf X}(0)$, one 
always finds the same evolved state ${\bf X}(t)$, at any later time $t>0$. Therefore, determinism
refers exclusively to ontic descriptions, and it does not deal with prediction. This has been clearly stressed 
by Atmanspacher (2002), in a paper by the rather eloquent title {\it Determinism is ontic, determinability is 
epistemic}. This distinction between ontic and epistemic descriptions was obvious to Maxwell; after
having noted the metaphysical nature of the problem of determinism in physics, he stated that:

{\it There are certain classes of phenomena ... in which a small error in the data only introduces a
small error in the result ... There are other classes of phenomena which are more complicated,
and in which cases of instability may occur.}

Also for Poincar\'e the distinction between determinism and prediction was rather clear, on the contrary, 
Popper (1992) confused determinism and prediction.

\section{Chaos and asymptotics}

Here, we briefly recall the essential properties of a deterministic chaotic system:
\\
I-  The evolution is given by a deterministic rule, for example, by a set of differential
equations;
\\
II- Solutions sensitively depend on the initial conditions: i.e., two initially almost identical states 
${\bf X}(0)$ and ${\bf X}'(0)$, with a very small initial displacement
$|{\bf X}'(0)- {\bf X}(0)|=\delta_0$, become separated at an exponential rate:
$$
|{\bf X}'(t)- {\bf X}(t)|=\delta_t \sim \delta_0 \, e^{\lambda t}\,,
\eqno(3)
$$
where $\lambda$ is positive and is called the Lyapunov exponent, for Bernoulli's shift $\lambda=\ln 2$;
\\
III- The evolution of the state ${\bf X}(t)$ is not periodic and appears quite irregular, similar in many 
respects to that of random systems.

The sensitive dependence on the initial condition drastically limits the possibility of 
making predictions: 
if the initial state is known with a certain uncertainty $\delta_0$, the evolution of 
the system can be 
accurately predicted with precision $\Delta$ only up to a time that depends on the Lyapunov exponent. This 
quantity is inherent in the system and does not depend on our ability to determine the initial state; hence, 
recalling Eq. (3), the time within which the error on the prediction does not exceed the desired
tolerance is:
$$
T_p \sim {1 \over \lambda} \ln {\Delta \over \delta_0}\,.
\eqno(4)
$$
The sensitivity to initial conditions introduces an error in predictions which grows 
exponentially in time. As the Lyapunov exponent $\lambda$ is an intrinsic characteristic of the system, predictions remain 
meaningful only within a time given by Eq. (4); therefore, it is well evident that a deterministic 
nature does not imply the possibility of an arbitrarily accurate prediction.

Let us note that, since ${\bf X}$ is in a bounded domain, some accuracy is needed in the definition
of the Lyapunov exponent: before one has to take the limit $\delta_0 \to 0$ and then $t \to \infty$:
$$
\lambda = \lim_{t \to \infty} \lim_{\delta_0 \to 0} {1 \over t} \ln \Big({ \delta_t \over \delta_0} \Big)\,.
$$
\\
\\
Another  important characterisation of the dynamics is given by the
Kolmogorov-Sinai entropy, $h_{KS}$, defined as follows. Just for the sake of simplicity we consider a 
system with discrete time:  
let ${\cal A}=\{ A_1, ... , A_N \}$ be a finite partition of the phase space (the space of configurations
of a given system under study), made up of the $N$ disjoint sets $A_i$, and consider the sequence of points
$$
\{ {\bf x}_1, ... , {\bf x}_n, ... \} \,,
$$
which constitutes the trajectory with initial condition ${\bf x}_0$. This trajectory can be associated with 
the symbol sequence
$$
\{ i_0, i_1, ..., i_n, ... \} \,,
\eqno(5)
$$
where $i_k=j$ if ${\bf x}_k \in A_j$.

Once a  partition ${\cal A}$ has been introduced, the coarse-grained properties of chaotic trajectories can be therefore studied 
through the discrete time sequence (5). Let $C_m=(i_1, i_2, ...  i_m)$ be a ``word'' (a string of symbols) of 
length $m$ and probability $P(C_m)$. The quantity
$$
H_m=\sum_{C_m} P(C_m) \ln P(C_m)
\eqno(6)
$$
is called the block entropy of the $m$-sequences\footnote{Shannon (1948) showed that, once the probabilities
$P(C_m)$ are known, the entropy (6) is the unique quantity which
measures, under natural conditions, the surprise or information carried by 
$\{ C_m \}$.}. In the limit of infinitely long sequences, the asymptotic entropy increment
$$
h_S({\cal A})=\lim_{m \to \infty} (H_{m+1}-H_m)
$$
is called the Shannon entropy, and in general depends on the partition ${\cal A}$ . Taking the largest
value over all possible partitions we obtain the so-called Kolmogorov-Sinai entropy:
$$
h_{KS}= \sup_{{\cal A}} h_S({\cal A})\,.
$$
A more intuitive definition of $h_{KS}$ starts from the partition ${\cal A}_{\epsilon}$
made of a grid of hypercubes with sides of length $\epsilon$, and takes the following limit:
$$
h_{KS}=\lim_{\epsilon \to 0} h(\epsilon)\,,
$$
where $h(\epsilon)=h_S({\cal A}_{\epsilon})$.
\\
\\
Naively, one might consider chaos in deterministic systems to be illusory, just a consequence of our 
observational limitations. Apparently, such a conclusion is confirmed by the fact that important measures 
of the dynamical complexity, such as the Lyapunov exponent $\lambda$ and the Kolmogorov-Sinai entropy 
$h_{KS}$, are defined via finite, albeit arbitrarily high, resolutions. For instance, in the computation 
of $\lambda$ one considers two trajectories, which are initially very close $|{\bf X}(0)- {\bf X}'(0)|=\delta_0$  
and diverge in time from each other. Similarly, $h_{KS}$ is computed introducing a partition of
the phase space, whose elementary cells have a finite size $\epsilon$. However, in the small-$\epsilon$
limit, $h(\epsilon)$ asymptotically tends to a value ($h_{KS}$) that no longer depends on $\epsilon$,
as happens to $\lambda$ in the small-$\delta_0$ limit. Therefore, $\lambda$ and $h_{KS}$ can be considered 
intrinsic properties of the dynamics themselves: they do not depend on our observational ability, provided it 
is finite, i.e., provided $\epsilon$ and $\delta_0$ do not vanish.
\\ 
According to Primas (2002), measures of stability, such as the Lyapunov exponent, concern ontic descriptions, 
whereas measures of information content or information loss, such as the Kolmogorov-Sinai entropy, relate to 
epistemic descriptions. We agree as far as stability is concerned. Regarding the epistemic character of $h_{KS}$, 
we observe that the Shannon entropy of a sequence of data, as well as the Kolmogorov-Sinai entropy, enjoy an 
epistemic status from a certain viewpoint, but not from another. The epistemic status arises from the fact 
that information theory deals with transmission and reception of data, which is necessarily finite. On the other 
hand, $h_{KS}$ is definitely an objective quantity, which does not depend on our observational limitations, 
as demonstrated by the fact that it can be expressed in terms of Lyapunov exponents 
(Cencini, {\it et al.} 2009). We note that the 
$\epsilon$-entropy $h(\epsilon)$ can be introduced even for stochastic processes, therefore it is a concept 
which links deterministic and stochastic descriptions.

\section{Chaos and probability}

After the (re)discovery of chaos in deterministic systems, owing to the presence of irregular and unpredictable 
behaviours, it is quite natural to adopt a probabilistic approach even in the deterministic realm.
Let us assume that we known the probability density of configurations in phase space at the initial time
$\rho({\bf x}, 0)$, it is possible to write down its time evolution law:
$$
\rho({\bf x}, 0) \to \rho({\bf x}, t) \,.
$$
Under certain conditions (mixing\footnote{The precise definition of mixing in dynamical systems requires 
several specifications and technicalities. To have an idea, imagine to put flour and sugar, in a given
proportion (say 40\% and 60\%, respectively) and initially separated, in a jar with a lid. After shaking the jar 
for a sufficiently long time, we expect the two components to be {\em mixed}, i.e., the probability to find flour or sugar 
in every part of the jar matches the initial proportion of the two components: a teaspoonful of the mixture
taken at random will contain 40\% of flour and 60\% of sugar.}) one has that al large time the probability 
density approaches a function which does not depend on $\rho({\bf x}, 0)$:
$$
\lim_{t \to \infty} \rho({\bf x}, t)=\rho^{inv}({\bf x})\,,\eqno(7)
$$
and is therefore called the invariant probability density. For instance, for Bernoulli's shift one has the 
following recursive rule:
$$
\rho(x, t+1)= {1\over 2} \rho\left({x \over 2 }, t\right)+{1 \over 2} \rho\left({x \over 2}+{1 \over 2}, t\right)\,,
$$
and the invariant probability density is constant in the interval $[0, 1]$:
$$
\lim_{t \to \infty} \rho(x, t)=\rho^{inv}(x)=1\,.
$$

It is rather natural, from an epistemic point of view, to accept the above probabilistic approach:
the introduction of $\rho({\bf x}, 0)$ can be seen as a necessity stemming from the human practical 
impossibility to determine the initial condition. For instance, in the case of Bernouilli's shift, knowing 
that the initial condition $x(0)$ is in interval $[x^{*}, x^{*}+\Delta]$, it is natural to assume that 
$\rho(x, 0)=1/\Delta$ for $x \in [x^{*}, x^{*}+\Delta]$, and $0$ otherwise.

For $t$ large enough (roughly $t > t_{*} \sim \ln (1/\Delta)$) one has the convergence of 
$\rho(x,t)$ toward the invariant probability distribution. Let us note that such a feature holds for any finite 
$\Delta$, while $t_{*}$ weakly depends on $\Delta$, therefore we can say that $\rho^{inv}(x)$, as well the approach 
to the invariant probability density, sure are objective properties independent of the uncertainty $\Delta$.
Perhaps somebody could claim that, since it is necessary to have $\Delta \ne 0$, the above properties, although 
objective, still have an epistemic character. We do not insist further.

Figs. 1 and 2 show, in a rather transparent way, how the approach to the $\rho^{inv}(x)$ 
is rather fast and basically independent on the  $\rho(x,0)$.
\begin{figure}
\includegraphics[width=7cm]{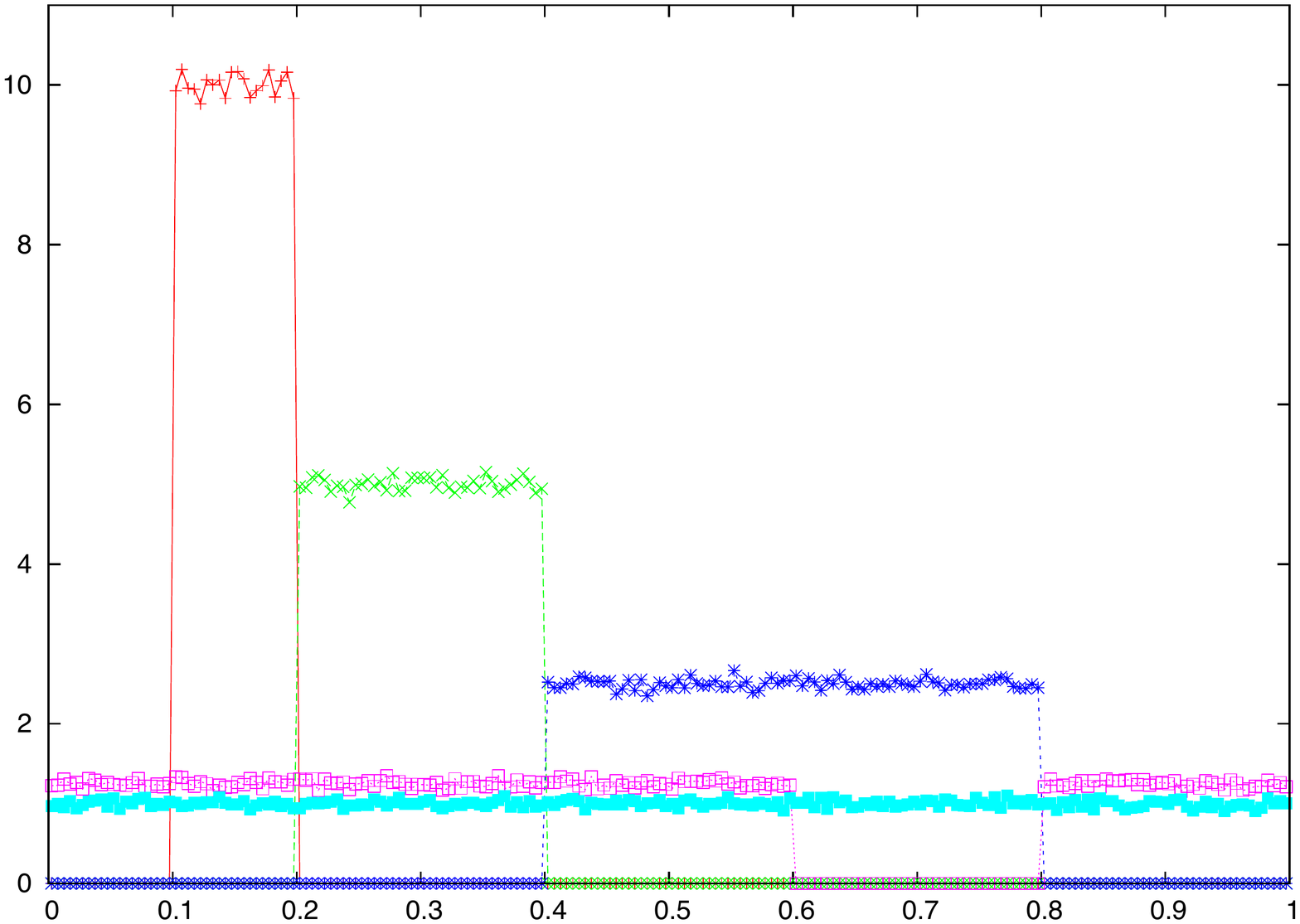}
\includegraphics[width=7cm]{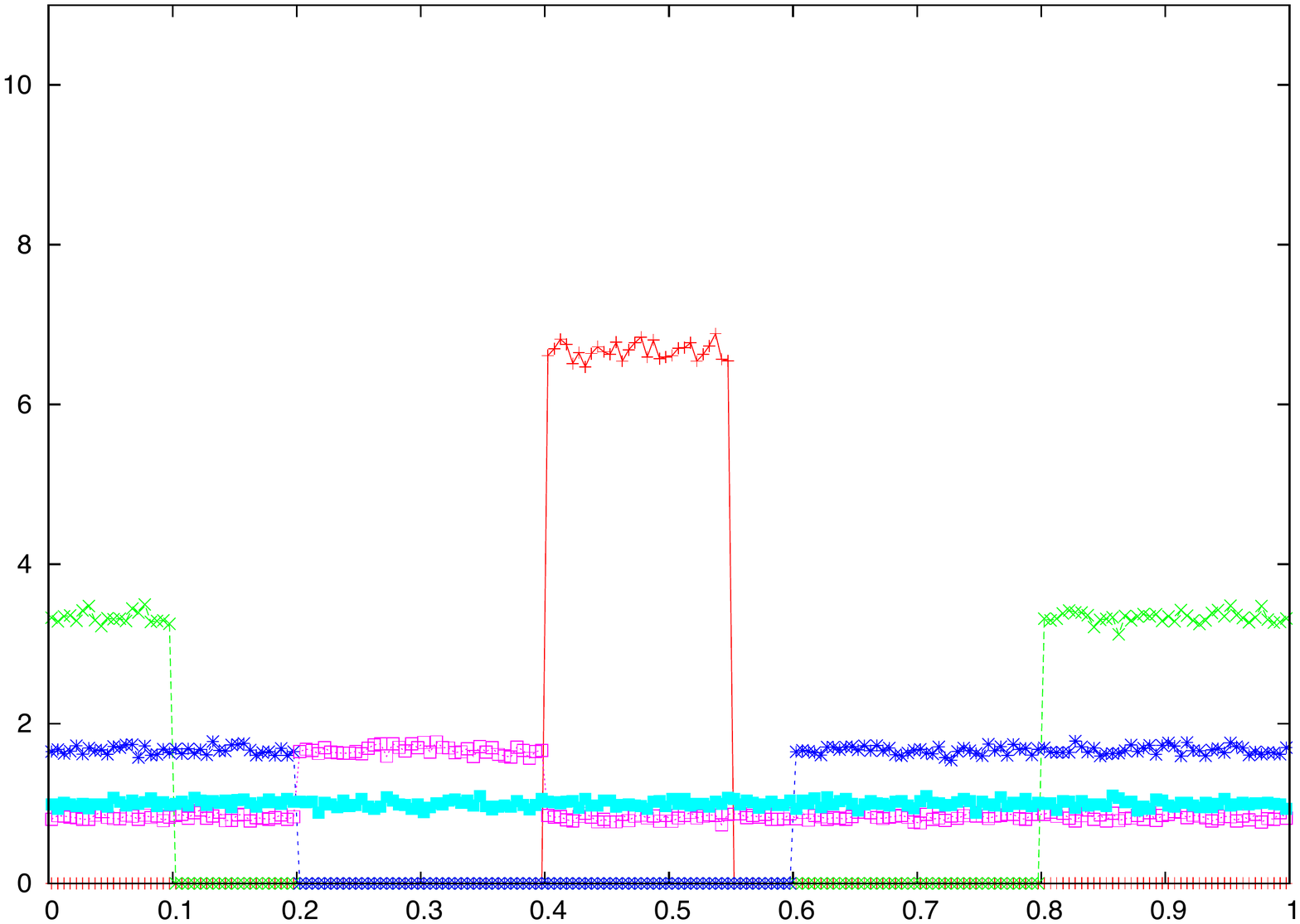}
\caption{\small{Probability density $\rho(x,t)$ for Bernoulli's shift. The distribution is obtained generating
at random 100000 points, uniformly distributed in the interval $[0.1:0.2]$ (a) and $[0.4:0.55]$ (b), and then iterating
the dynamics of each point for 14 time steps. The red curves correspond to the initial distribution, the green, the
blue, and the magenta curves correspond to one, two and three time steps, respectively. As it is evident, at each time step 
the dynamics initially doubles the width of interval over which the points are distributed, until 
$\rho(x,t)\approx\rho^{inv}(x)=1$. The light blue curves are obtained after 14 iterations: The probability density
$\rho(x,t)$ for $t> 12$ is close to the invariant distribution $\rho^{inv}(x)$ independently of the initial 
$\rho(x,0)$.}}
\end{figure}

We saw how chaotic systems and, more precisely, those which are ergodic\footnote{A very broad definition of 
an ergodic system relies on the identification of time averages and averages computed with the
invariant probability density (7). Said in other
words, a system is ergodic if its trajectory in phase space, during its time evolution, visits (and densely explores) 
all the accessible regions of phase space, so that the time spent in each region is proportional to the invariant
probability density assigned to that region.
Therefore, if a system is ergodic, one can understand its statistical features looking at the
time evolution for a sufficient long time; the conceptual and technical relevance of ergodicity
is quite clear.},
 naturally lead to 
probabilistic descriptions, even in the presence of deterministic dynamics. In particular, ergodic theory justifies 
the frequentist interpretation of probability, according to which the probability of a given event is defined 
by its relative frequency. Therefore, assuming ergodicity, it is possible to obtain an empirical notion of 
probability which is an objective property of the trajectory (von Plato 1994). There is no universal agreement 
on this issue; for instance, Popper (2002) believed that probabilistic concepts are extraneous to a 
deterministic description of the world, while Einstein held the opposite view, as expressed in his letter to 
Popper:

{\it I do not believe that you are right in your thesis that it is impossible to derive statistical
conclusions from a deterministic theory. Only think of classical statistical mechanics (gas
theory, or the theory of Brownian movement)}.

\section{A brief digression on models in physics}

At this point of the discussion, we wish to recall that our description of physical
phenomena is necessarely based upon models\footnote{There are several definitions of a {\em Model}, but to our
purposes the following is a reasonable one: Given any system $\mathcal S$, by which we mean a set of objects
connected by certain relations, the system $\mathcal M$ is said a model of $\mathcal S$ if a correspondence
can be established between the elements (and the relations) of $\mathcal M$ and the elements (and the relations) 
of $\mathcal S$, by means of which the study of $\mathcal S$ is reduced to the study of $\mathcal M$, within 
certain limitations to be specified or determined.},
that entail a schematization of a specified portion of the physical world. Take for instance
the pendulum described by Eq. (1). The mathematical object introduced thereby relates to a physical pendulum
under some specific assumptions. For instance, the string of length $L$, that connects the swinging body to 
a suspension point, is assumed to be {\em inextensible}, whereas any 
physical string is (to some extent) extensible. The model also assumes that gravity is spatially
uniform and does not change with time, i.e., it can be described by a constant $g$. Eq. (1) will 
therefore reasonably describe a physical pendulum only inasmuch as the variations of $L$
and/or $g$ are sufficiently small, so as to add only tiny corrections. Even more important,
there is not in the physical world such an object as an isolated pendulum, whereas Eq. (1) totally ignores
the physical world around the pendulum (the only ingredients being gravity, the string, and the suspension point). 
Galileo Galilei was well aware of this subtlety when comparing the prediction of our mathematical models with the 
physical phenomena they aim to describe\footnote{Experiments are usually carried out under {\em controlled conditions},
meaning that every possible care is taken in order to exclude external influences and focus on specific aspects of the 
physical world. In his ``Dialogues concerning 
two new sciences'', Galilei (English translation, 1914) describes the special care to be taken in order to keep the
accidents under control: ``... I have attempted 
in the following manner to assure myself that the acceleration actually experienced by falling bodies is that above 
described. A piece of wooden moulding or scantling, about 12 cubits long, half a cubit wide, and three finger-breadths 
thick, was taken; on its edge was cut a channel a little more than one finger in breadth; having made this groove 
{\em very straight, smooth, and polished}, and {\em having lined it with parchment}, also {\em as 
smooth and polished as possible}, we rolled along it a {\em hard, smooth, and very round} bronze ball ...'' (the 
italicized emphases are ours).} and called {\em accidents} (on this topic, see, e.g., Koertge, 1977) all external 
influences apt to modify, often in an apparently unpredictable way, the behaviour of a (supposedly isolated) portion of 
the physical world. In the case of the Eq. (1), we are, e.g., neglecting the fact that a real pendulum swings in 
a viscous medium (the air), and also experiences some friction at the suspension point. These effects gradually alter 
the motion of the pendulum, which is no longer periodic and eventually stops. Eq. (1) also neglects the
fact that the Earth is not an inertial reference frame: it rotates around its axis and around the Sun.
The first effect is far more important and gives rise to the gradual but sizable variation of the plane of 
oscillation (Foucault's pendulum). There are several external influences that may alter the motion of a pendulum.
Some of them may be accounted for, at least to some extent, by simple modifications of Eq. (1). Other
are rather complicated and are not easily accountable. Thus, Eq. (1) describes a pendulum only as far and as 
long as external influences do not alter significantly its motion. Said in other words, it describes a pendulum 
under {\em controlled conditions}.

\section{The old dilemma determinism/stochasticity}

The above premise underlines the crucial importance of the concept of {\it state of the system}, i.e., in 
mathematical terms, the variables which describe the phenomenon under investigation. The relevance of 
such an aspect is often underestimated; only in few situations, e.g., in mechanical systems, it is easy to 
identify the variables which describe the phenomenon. On the contrary, in a generic case, there are serious 
difficulties; we can say that often the main effort in building a theory of nontrivial phenomena concerns 
the identification of the appropriate variables. Such a difficulty is well known in statistical physics; it 
has been stressed, e.g., by Onsager and Machlup (1953) in their seminal work on fluctuations and irreversible 
processes, with the caveat:

{\it how do you know you have taken enough variables, for it to be Markovian?}

In a similar way, Ma (1985) notes that:

{\it the hidden worry of thermodynamics is: we do not know how many coordinates
or forces are necessary to completely specify an equilibrium state.}

Unfortunately, we have no definite method for selecting the proper variables.

Takens (1981) showed that from the study of a time series $\{ u_1, u_2, ..., u_m \}$, where $u$ is an observable sampled 
at the discrete times $t_j= j \Delta t$ and $u_j=u(t_j)$, it is possible (if we know that the system is 
deterministic and is described by a finite dimensional vector) to determine the proper variable ${\bf X}$.
Unfortunately the method has rather severe limitations:
\\
a) It works only if we know a priori that the system is deterministic;
\\
b) The protocol fails if the dimension of the attractor\footnote{The attractor of a dynamical system is 
a manifold in phase space toward which the system tends to evolve, regardless of the initial conditions.
Once close enough to the attractor, the trajectory remains close to it even in the presence of small 
perturbations.} is large enough (say more than $5$ or $6$). 
\\
Therefore the method cannot be used, apart for special cases (with a small dimension), to build up a model from 
the data.

We already considered arguments, e.g., by van Kampen, which deny that determinism may be decided on the basis 
of observations. This conclusion is also reached from detailed analyses of sequences of data produced by the 
time evolutions of interest. In few words: the distinction between deterministic chaotic systems and genuine 
stochastic processes is possible if one is able to reach arbitrary precision on the state of the system.

Computing the so-called $\epsilon$-entropy $h(\epsilon)$, at different resolution scales $\epsilon$, at least 
in principle, one can distinguish potentially underlying deterministic dynamics from stochastic ones. 

From a mathematical point of view the scenario is quite simple:
for a deterministic chaotic system as $\epsilon \to 0$ one has $h(\epsilon) \to h_{KS} <\infty$,
while for stochastic processes $h(\epsilon) \to \infty$\footnote{Typically $h(\epsilon) \sim \epsilon^{-\alpha}$
where the value of $\alpha$ depends on the process under investigation (Cencini, {\em et al.}, 2009).}.
On the other hand an arbitrary solution is not possible, therefore the analysis of temporal series can
only be used, at best, to pragmatically classify the stochastic or chaotic character of the observed signal, 
within certain scales (Cencini, {\em et al.}, 2009; Franceschelli, 2012). At first, this could be disturbing: not even the most 
sophisticated time-series analysis that we could perform reveals the ``true nature'' of the system under 
investigation, the reason simply being the unavoidable finiteness of the resolution we can achieve.

On the other hand, one may be satisfied with a non-metaphysical point of view, in which the true nature of 
the object under investigation is not at stake. The advantage is that one may choose whatever model is more 
appropriate or convenient to describe the phenomenon of interest, especially considering the fact that, 
in practice, one observes (and wishes to account for) only a limited set of coarse-grained properties. 
\\
In light of our arguments, it seems fair to claim that the vexed question of whether the laws of physics are 
deterministic or probabilistic has, and will have, no definitive answer. On the sole basis of empirical 
observations, it does not appear possible to decide between these two contrasting arguments:
\\
(i)  Laws governing the Universe are inherently random, and the determinism that is
believed to be observed is in fact a result of the probabilistic nature implied by
the large number of degrees of freedom;
\\
(ii) The fundamental laws are deterministic, and seemingly random phenomena appear so due to deterministic 
chaos.

Basically these two positions can be viewed as a reformulation of the endless debate on quantum mechanics: 
thesis (i) expresses the inherent indeterminacy claimed by the Copenhagen school, whereas thesis (ii) illustrates 
the hidden determinism advocated by Einstein (Pais 2005).
\\
\\
\\
\\
{\bf References}
\\
\\
* Atmanspacher, H.
``Determinism is ontic, determinability is epistemic''.
 In: Atmanspacher, H., Bishop, R. (eds.) 
 {\it Between Chance and Choice}.
 (Imprint Academic, Thorverton, 2002)
\\
* Boyd, R.
``Determinism, laws and predictability in principle''
Phylosophy of Science {\bf  39}, 43 (1972)
\\
* Campbell, L., Garnett, W.
{\it  The Life of James Clerk Maxwell}
(MacMillan and Co., London, 1882)
\\
* Cencini, M., Cecconi, F., Vulpiani, A.
{\it  Chaos: From Simple Models to Complex Systems}
(World Scientific, Singapore, 2009)
\\
* Chibbaro, S., Rondoni, L., Vulpiani, A.
{\it Reductionism, Emergence and Levels of Reality}
(Springer-Verlag, Berlin, 2014)
\\
* Dyson, F.
``Birds and frogs''
 Not. AMS {\bf 56}, 212 (2009)
\\
* Franceschelli, S. ``Some remarks on the compatibility between determinism and unpredictability''
Progr. in Bioph. and Mol. Bio. {\bf 110}, 61 (2012)
\\
* Galilei, G. ``Dialogues concerning two new sciences'' (MacMillan, New York, 1914)
\\
* van Kampen, N. G.
``Determinism and predictability''
 Synthese {\bf 89}, 273 (1991)
\\
* Koertge, N. ``Galileo and the problem of accidents'', Journal of the History of Ideas {\bf 38}, 389 (1977).
\\
* Lorenz, E. N.
``Deterministic nonperiodic flow''
 J. Atmos. Sci {\bf 20}, 130 (1963)
 \\
* Ma, S. K.
 {\it  Statistical Mechanics}
 (World Scientific, Singapore, 1985)
\\
* Onsager, L., Machlup, S.
``Fluctuations and irreversible processes''
 Phys. Rev. {\bf 91}, 1505 (1953)
\\
* Pais, A.
{\it Subtle is the Lord: The Science and the Life of Albert Einstein}
(Oxford University Press, 2005)
\\
* Poincar\'e, H.
{\it  Les m\'ethodes nouvelles de la m\'ecanique c\'eleste}
(Gauthier-Villars, Paris, 1982)
\\
* Popper, K.R.
{\it The Open Universe: An Argument for Indeterminism. From the Postscript to the
Logic of Scientific Discovery}
(Routledge, London, 1992)
\\
* Popper, K. R.
{\it The Logic of Scientific Discovery}
(Routledge, London, 2002)
\\
* Prigogine, I., Stengers, I.
{\it Les Lois du Chaos}
 (Flammarion, Paris 1994)
 \\
* Primas, H.
``Hidden determinism, probability, and times arrow''
 In: Bishop, R., Atmanspacher, H. (eds.) 
 {\it Between Chance and Choice}, p. 89
 (Imprint Academic, Exeter 2002)
\\
* Shannon, C. E.
``A note on the concept of entropy''
 Bell System Tech. J. {\bf 27}, 379 (1948)
\\
* Stone, M. A.
"Chaos, prediction and Laplacean determinism"
Am. Phylos. Quarterly {\bf 26}, 123 (1989)
\\
*  Takens, F.
 "Detecting strange attractors in turbulence", In: D. Rand, L.-S. Young (eds.),
 {\it  Dynamical Systems and Turbulence},  Lecture Notes in Mathematics {\bf 898} (1981),
Springer-Verlag.
\\
* von Plato, J.
{\it Creating Modern Probability}
(Cambridge University Press, 1994)
\\
* Vauclair, S.
{\it El\'ementes de Physique Statistique}
(Inter\'editions, Paris 1993)
 \end{document}